\documentclass[prd, aps, nofootinbib, preprintnumbers, showpacs, superscriptaddress, twocolumn, floatfix]{revtex4}

\usepackage{amssymb}
\usepackage[english]{babel}
\usepackage{ifpdf}
\usepackage[latin9]{inputenc}
\usepackage{mathpazo}
\usepackage{mathrsfs}
\usepackage{float}
\usepackage{psfrag}
\usepackage{color}
\usepackage{hyperref}

\ifpdf
\usepackage{hyperref}
\fi

\usepackage{graphicx}

\newcommand{\be}{\begin{equation}}
\newcommand{\ee}{\end{equation}}

\newcommand{\bee}{\begin{eqnarray}}
\newcommand{\eee}{\end{eqnarray}}

\usepackage{soul}

\begin{document}

\title{Are gravitational waves from giant magnetar flares observable?}


\author{Burkhard Zink} 
\author{Paul D. Lasky}
\author{Kostas D. Kokkotas}
\affiliation{Theoretical Astrophysics, Institute for Astronomy and Astrophysics, University of T\"ubingen, Auf der Morgenstelle 10, 
    T\"ubingen 72076, Germany}

\begin{abstract}

Are giant flares in magnetars viable sources of gravitational radiation? 
Few theoretical studies have been concerned with this problem, with the small number using either highly idealized
models or assuming a magnetic field orders of magnitude beyond what is supported by observations.
We perform nonlinear general-relativistic magnetohydrodynamics
simulations of large-scale hydromagnetic instabilities in magnetar
models. We utilise these models to find gravitational wave emissions over a wide range of energies, from $10^{40}$ to $10^{47} \, \mathrm{erg}$. This allows
us to derive a systematic relationship between the surface field strength and the gravitational wave strain, which
we find to be highly nonlinear. In particular, for typical magnetar fields of a few times $10^{15} \, \mathrm{G}$, 
we conclude that a direct observation of $f$-modes excited by global magnetic field reconfigurations is unlikely with
present or near-future gravitational wave observatories, though we also discuss the possibility that modes in
a low-frequency band up to $100 \, \mathrm{Hz}$ could be sufficiently excited to be relevant for observation.

\end{abstract}

\maketitle

\emph{Introduction.} Bursts and occasionally giant flares in
soft-gamma repeaters (SGRs) and anomalous X-ray pulsars (AXPs) are
commonly understood as consequences of magnetic field activity
in magnetars, neutron stars endowed with strong magnetic fields \cite{Mereghetti2008}. 
These violent events have been considered in the literature as possible efficient sources of gravitational
radiation, since they are very compact, and since the high
electromagnetic luminosity may represent only a small part of the overall
energy contained in the mechanism. It is thus not surprising that operational gravitational wave observatories,
in particular LIGO and VIRGO, have been employed to establish upper limits on 
SGR bursts and storms. Presently, the best empirically derived upper limits are $1.4 \times 10^{49} \mathrm{erg}$
for the $f$-mode, and $3.5 \times 10^{44} \, \mathrm{erg}$ for a white noise band around 
$100 \, - \, 200 \, \mathrm{Hz}$ \cite{Abadie2011}. For the $f$-mode, gravitational wave emission of this magnitude
would imply a very substantial excitation of the star. Can luminosities of this order actually be
realized in a nearby giant flare?

There has been little theoretical work addressing this question, for a large part because the system
is very complex and the giant flare mechanism is not yet well
understood. Possible flare mechanisms are discussed in \cite{Thompson1995, Thompson2001, Lyutikov2006, Gill2010, Levin2011}
and fall into two broad categories: internal
mechanisms based on a large-scale rearrangement of the interior magnetic field, and external mechanisms 
likely following magnetic reconnection in the crust and magnetosphere. In this work, we will assume
a scenario involving a global restructuring of the magnetic field.

Two direct pieces of evidence correlate the electromagnetic signal
from magnetar flares with their magnetic field 
dynamics -- energy budgets associated with the flare luminosities \cite{Thompson1995} and observations 
of quasi-periodic oscillations in the tails of the flares \cite{Israel2005, Colaiuda2011, Gabler2011}.  
An {\it optimistic} estimate for gravitational wave emissions from magnetic field dynamics can therefore be attained by assuming a large-scale, dynamical rearrangement of the core magnetic field inside the star.  
Ioka \cite{Ioka2001} has investigated the maximum gravitational wave energy released by the change in moment of inertia induced by
 such a mechanism and placed an upper limit 
of about $10^{49} \mathrm{erg}$ under ideal conditions, including optimistic values of the internal magnetic field. 
More recently, Corsi and Owen \cite{Corsi2011} found similar
values to be possible under more generic conditions, still tapping into the full energy reservoir associated with an instantaneous change in the magnetic potential energy of the star. In contrast, Levin and van Hoven \cite{Levin2011} do not find $f$-mode
detection to be very likely in the near future.  Their model is based on the aforementioned external mechanism triggering $f$-modes in the star.  In the present paper, we focus on the other side of the problem -- attempting to trigger large-scale mass motions that could generate gravitational radiation through a global rearrangement of the \emph{internal} magnetic field.

Our recent publication on dynamical instabilities \cite{Lasky2011} has focused 
on the dynamics and evolution of magnetic fields inside relativistic stellar models, and we have followed the development of
these instabilities until saturation and ring-down to establish the nature of the quasi-stationary states which result from these
processes. Most recently, Ciolfi et al \cite{Ciolfi2011} have performed numerical simulations
of the same hydromagnetic instabilities, and concluded that giant flares \emph{could} give rise to observable
gravitational radiation, employing a stellar model with a surface field strength of $6.5 \times 10^{16} \, \mathrm{G}$ 
to reach this conclusion. Actual SGR field strengths are closer to $1/60$ of this value \cite{Mereghetti2008, url:McGillWeb}, 
and the luminosity depends on the field strength in a 
highly nonlinear fashion, as we shall see below.

In this paper, we will address the following questions: (i) Assuming a large-scale rearrangement of the core
magnetic field to be involved in a giant flare, how does the gravitational wave luminosity scale with the field strengths? 
(ii) As a consequence of this relation, can we expect to observe magnetar giant flares in gravitational wave detectors? 

% ======= Model =============

\emph{Model.} In order to investigate gravitational wave luminosities, we consider a
large-scale restructuring of an initially purely poloidal magnetic field. While an actual
magnetar should be quasi-stationary before the flaring event, we employ a hydromagnetically 
unstable star to investigate the gravitational wave signal associated with the instability.
This provides us with a simplified model that mimics a catastrophic reconfiguring of the internal magnetic 
field during a magnetar flare.

As with our recent publication \cite{Lasky2011}, we are using the graphics processing unit (GPU)-accelerated \textsc{Horizon}
code \cite{Zink2011Horizon} for general relativistic magnetohydrodynamics in the Cowling approximation, 
in conjunction with the \textsc{Lorene} library for
the construction of magnetized equilibrium neutron stars \cite{Bocquet1995}. \textsc{Horizon} is based on the \textsc{Thor} code
\cite{Zink:2010, Korobkin:2011}, but employs GPUs for high-throughput parallel processing. 
Gravitational waves are extracted using the first-moment form of the quadrupole formula.  
For more details on the numerical method and stellar models see \citet{Zink2011Horizon, Lasky2011}.

Our initial models are relativistic equilibrium $n = 1$ polytropes
(mass $1.3 \, M_\odot$, radius $15 \ \mbox{km}$) with purely poloidal magnetic fields of varying strength. 
In particular, we have investigated a sequence of 
models with polar \emph{surface} magnetic field amplitudes between $B_{pole}=3.1
\times 10^{15} \, \mathrm{G}$ and $5.5 \times 10^{16} \, \mathrm{G}$ corresponding to magnetic field 
strengths of $1.6 \times 10^{16} \, \mathrm{G}$ to $2.7 \times 10^{17} \, \mathrm{G}$ in the center
of the star. While typical magnetar field strengths reside near the low end
of this spectrum, we have decided on a sequence of models to gain
fundamental insights into the relation of initial field configuration and the gravitational wave amplitude.

Purely poloidal fields are known to be dynamically unstable
\cite{Braithwaite2007, Lasky2011}, and we follow the evolution 
throughout development, saturation \emph{and} ring-down of the instability. This is particularly
challenging for models with lower field strengths, since we need to follow the system for many Alfv\'en times in all cases.

% ======= Results =============

\begin{figure}[htb]
\includegraphics[width=\columnwidth]{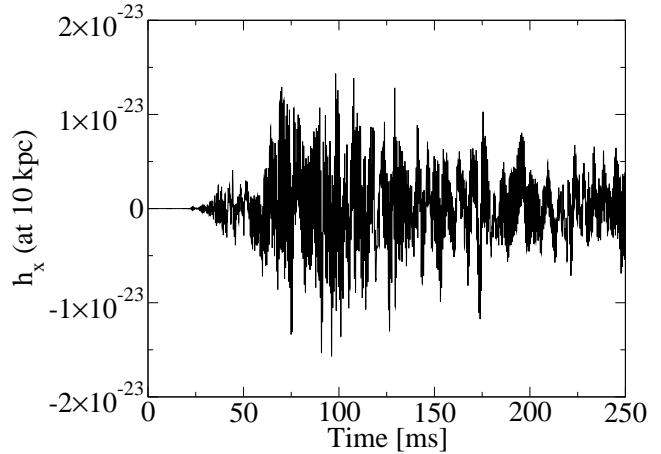}
\caption{
Gravitational wave signal obtained from one of our simulations. This particular stellar model has an initial
surface magnetic field strength of $1.8 \times 10^{16} \, \mathrm{G}$ and develops a poloidal field
instability during the first $50 \ \mathrm{ms}$ (see also \cite{Lasky2011}) which restructures the 
global magnetic field inside the star. 
The gravitational wave signal, here $h_\times$ for an assumed 
source distance of $10 \, \mathrm{kpc}$, has the expected $1.8 \, \mathrm{kHz}$ oscillations associated
with the $f$-mode, and also exhibits low frequency components which are further discussed in the text.
}
\label{fig:h_e_cross_1e17}
\end{figure}

\emph{Results.} All our models exhibit the poloidal field instability as expected. In figure \ref{fig:h_e_cross_1e17}, we show the measured 
gravitational wave strain $h_\times$ at $10 \, \mathrm{kpc}$ in a particular model with $B_{pole}=1.8\times10^{16}\,\mathrm{G}$. Other components and
other models in the sequence show a similar structure, although the growth timescale of the instability and the
strain amplitude are different. An analysis of the signal spectrum shows that the fundamental quadrupole $f$-mode is excited by the magnetic
field instability. Intriguingly, we also see some evidence that \emph{low-frequency} modes in the range of $100 \, \mathrm{Hz}$ may contribute to the gravitational wave signal.
This evidence is not yet conclusive and warrants further investigation, in particular
very long-term simulations in the order of a second or more to gain a sufficiently high resolution in this part of the spectrum. 

\begin{figure}[htb]
\includegraphics[width=\columnwidth]{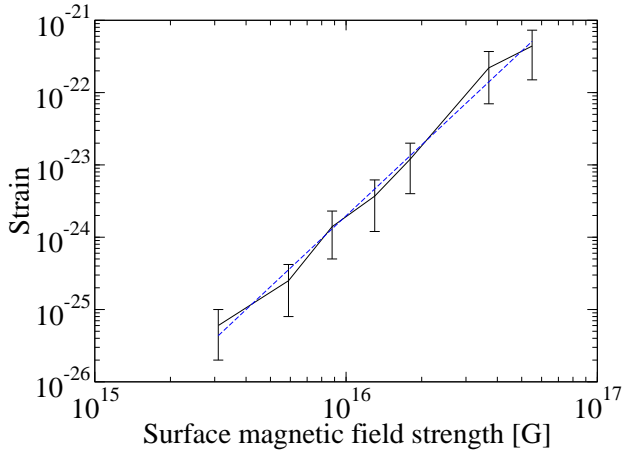}
\caption{Gravitational wave strain in relation to surface magnetic field
strength at the stellar pole. This plot represent results from our entire sequence of simulations,
with surface magnetic fields ranging from  $3.1\times 10^{15} \, \mathrm{G}$ to  $5.5 \times 10^{16} \, \mathrm{G}$.
The signal amplitude is found to depend strongly on the magnetic field strength, ranging from almost $10^{-21}$ for
our most extreme model down to less than $10^{-25}$ for models closer to realistic magnetar strengths. The dashed 
line is a power-law fit to the data (see text).}
\label{fig:strain_surfB}
\end{figure}

Figure \ref{fig:strain_surfB} collects the gravitational wave amplitudes as a function of the magnetic field strength. The
approximate values and error bars for the strains are derived from the time variation in the $h_\times$ signals 
(which are similar in amplitude to $h_+$) after the instability has saturated.
We find an approximate power-law relation between $h$ and $B_{pole}$ given by

\begin{eqnarray}
h \approx 1.1 \times 10^{-27} \times \left( \frac{10 \, \mathrm{kpc}}{d} \right)  \times \left( \frac{B_{pole}}{10^{15} \, \mathrm{G}} \right)^{3.3}.
\label{eq:hmax}
\end{eqnarray}

Most of the energy in the signal is in the $f$-mode. Assuming a gravitational wave damping time of approximately 
$100 \, \mathrm{ms}$ (e.g. \cite{Lindblom1983, Andersson98b}), we find a corresponding power-law relation for the energy
 emitted in gravitational radiation:

\begin{eqnarray}
E_{gw} \approx 1.5 \times 10^{36} \times \left( \frac{B_{pole}}{10^{15} \, \mathrm{G}} \right)^{6.5} 
\, \mathrm{erg}.
\label{eq:egw}
\end{eqnarray}

From this result, which can be directly compared with observations \cite{Abadie2011}, we make three notes:
(i) The ratio of gravitational wave energy extracted from our
simulations to the electromagnetic energy
emitted in a flare \cite{Mereghetti2008} is low, in the order of
$E_{gw} / E_{em} \approx 10^{38} \, \mbox{erg} / 10^{44} \, \mbox{erg} = 10^{-6}$.
(ii) The gravitational wave amplitude is a highly nonlinear function of the surface
magnetic field strength. 
(iii) Typical magnetar field strengths
of $10^{15} \, \mathrm{G}$ give rise to strains \emph{below} $10^{-25}$ and energies lower than $10^{40} \, \mathrm{erg}$
for a source at $10 \, \mathrm{kpc}$, even if we assume a catastrophic global restructuring of the field to be associated with a giant flare.

\begin{figure}
\includegraphics[width=\columnwidth]{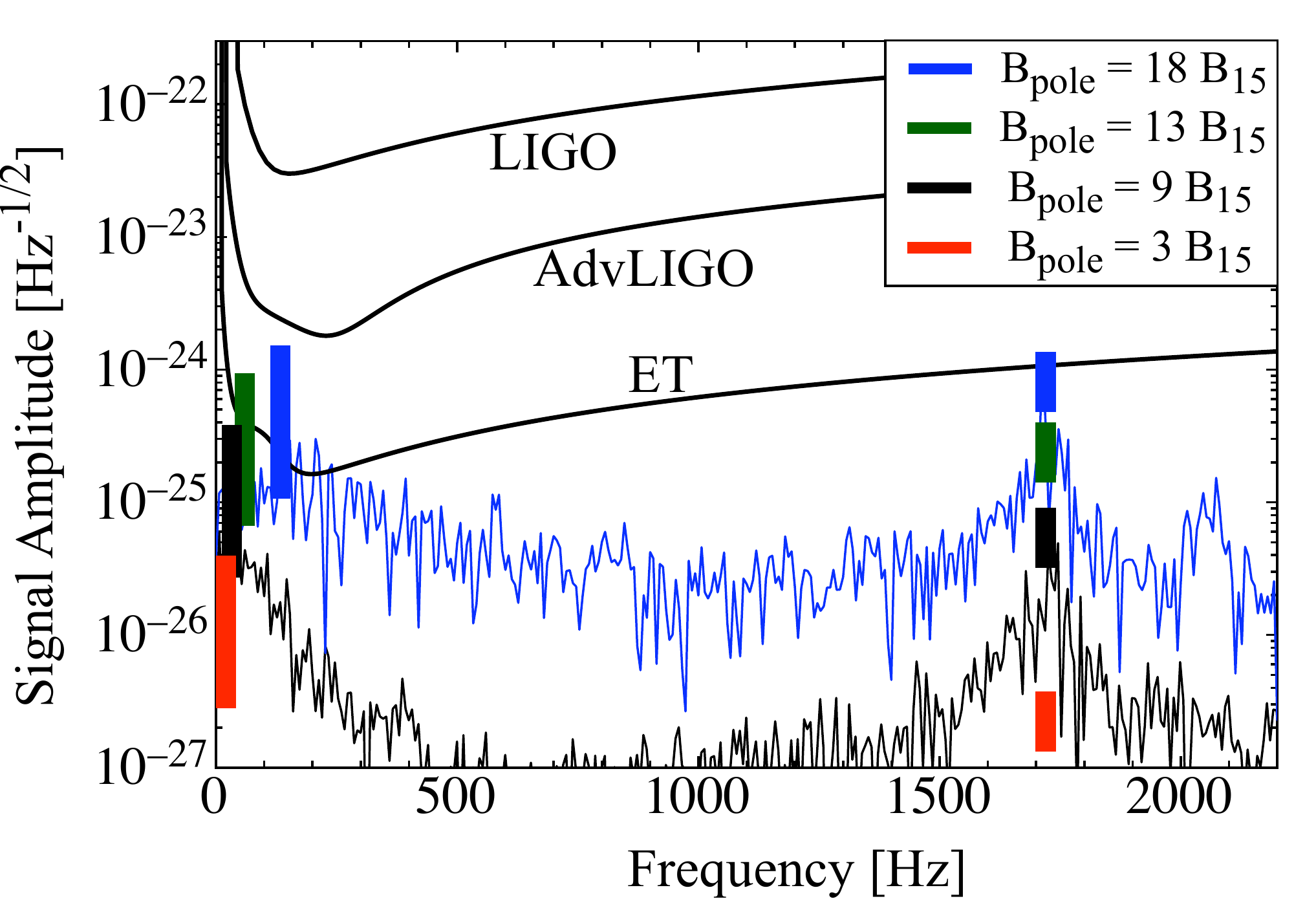}
\caption{Signal amplitude $\sqrt{T}\left|\tilde{h}(f)\right|$ against the oscillation frequency.  The colored boxes on the right represent the maximum for the 
$f$-mode assuming a constant periodic source lasting between 50 and 200 ms.  The colored boxes on the left represent the maximum
 mode seen in the Fourier transform for any given frequency below the $f$-mode frequency assuming a constant periodic source lasting 
between 10 ms and 1 s.  These scale with $\sqrt{T}$, implying one can easily extrapolate to alternative values of the damping time.  
We have further plotted the entire spectrum (assuming a damping time of $T=100\,\mbox{ms}$) for both the $B_{pole}=1.8\times10^{16}\,\mbox{G}$ 
model (blue line) and the $B_{pole}=8.8\times10^{15}\,\mbox{G}$ model
(black line). The noise power spectral density curves have been taken from the review
 article of \citet{sathyaprakash09}.}
\label{fig:h_e_cross_gw_SNR}
\end{figure}

We have indicated the signal-to-noise ratios we derive from our data for different detector sensitivity
curves in figure~\ref{fig:h_e_cross_gw_SNR}.  Here we have plotted the signal amplitude $\sqrt{T}\left|\tilde{h}(f)\right|$, where $T$ is the damping time of the oscillation and $\tilde{h}(f)$ is the Fourier transform of $h_{\times}$, as a function of the frequency for various magnetic field strength models.  At approximately $1.8\,\mathrm{kHz}$ we show the signal amplitude for the $f$-mode as excited by the hydromagnetic instability and assuming a damping time of $50\,\mathrm{ms}\le T\le200\,\mathrm{ms}$.  In the lower part of the spectrum we plot the other maximal mode seen in the Fourier transform, assuming a damping time between of $10\,\mathrm{ms}\le T\le 1\,\mathrm{s}$.  Over this we plot the entire spectrum (assuming $T=100\,\mathrm{ms}$) 
for two models with $B_{pole}=8.8 \times 10^{15}\,\mbox{G}$ and $B_{pole}=1.8\times 10^{16}\,\mbox{G}$ respectively.  Finally, we also plot the root of the noise power spectral density, $\sqrt{\left|S_{h}(f)\right|}$, as a function of frequency for the LIGO, AdvLIGO and ET detectors
  \cite{sathyaprakash09}.

The amplitude signal-to-noise ratio, defined by $\sqrt{T}\left|\tilde{h}(f)\right|/\left|S_{h}(f)\right|^{1/2}$ can 
be read off the graph as the ratio between the signal and the noise curve for the respective detector.
The main conclusion from this figure is this: \emph{Assuming that
a giant flare is associated with a catastrophic large-scale rearrangement of the core magnetic field, 
the gravitational wave signal associated with $f$-modes are not observable with present or near-future gravitational wave
observatories.} 

This statement is in line with the conclusion of \cite{Levin2011} concerning the efficiency of the internal mechanism
in exciting $f$-mode radiation.  Moreover, it is in contrast with the scenarios presented in \cite{Ioka2001, Corsi2011}, which
are both based on equilibrium models as opposed to full dynamical simulations. While we cannot exclude that different
initial model configurations could produce higher levels of $f$-mode radiation, we consider it, at present, unlikely 
that the result will change by many orders of magnitude.

Our results can also be compared with the numerical study of \cite{Ciolfi2011} which obtains a much more optimistic 
prediction than we do, although they investigate the same kind of instability. However, the magnetar model 
considered by these authors has a very strong surface 
magnetic field of $6.5 \times 10^{16} \, \mathrm{G}$, which is even stronger than the most extreme case we
have considered here. Such a high magnetic field strength favors a coupling of Alfv\'en modes into
$f$-modes because the Alfv\'en speed approaches the speed of sound. 
Clearly, the observed soft-gamma repeaters do not fall into this category.

% ========== Discussion ============

\emph{Discussion.} We have performed three-dimensional general-relativistic magnetohydrodynamics simulations of neutron stars,
 inducing large-scale reconfigurations of the internal magnetic field to model gravitational wave emissions from magnetar flares.
  We have found power-law relations governing the surface, polar magnetic field strength as a function of the maximal
 gravitational wave strain (equation \ref{eq:hmax}) and also the energy emitted in gravitational waves (equation \ref{eq:egw}).  
We find that the gravitational wave emissions due to $f$-mode excitations are unlikely to be observed in current or near-future 
gravitational wave observatories (figure \ref{fig:h_e_cross_gw_SNR}).

There are two factors which could modify our above conclusion. The first is that other equations of state and initial magnetic field topologies
could be investigated, e.g. those with a dominating toroidal field component. Mixtures of poloidal and
toroidal fields in neutron stars may be stable for a range of energy ratios \cite{Braithwaite2009} and could indeed represent
typical field equilibria. However, there is a possibility that configurations with a strong toroidal component (and comparatively \emph{weak}
surface dipole field) could be unstable and lead to higher coupling into the $f$-modes and consequently stronger
gravitational wave luminosities, while still being consistent with SGR spindown rates. Recent studies of hydromagnetic equilibria
in stratified neutron stars \cite{Glampedakis2011, Lander2011}, however, cast a shadow of doubt over this possibility.

The other factor is the possible low-frequency component of the signal (see also \cite{Levin2011, Kashiyama2011}). 
As discussed above, we have found indications (but no solid confirmation) of spectral components in the band typically 
associated with g-modes or Alfv\'en modes, i.e. below 200 Hz.
This is particularly interesting because the detectors are most sensitive in this regime. Moreover, these modes
have much longer damping times and are assumed to be part of the post-flare signal.
We leave these possibilities for future work. 

\acknowledgments{We would like to thank R. Ciolfi, K. Glampedakis,
S. Lander and L. Rezzolla for helpful discussions. This work was supported by the SFB/Transregio 7 on ``Gravitational Wave Astronomy''
by the DFG. Some computations were performed on the Multi-modal Australian ScienceS Imaging and Visualisation
Environment \MakeUppercase{MASSIVE} (\url{www.massive.org.au}) and the GPU nodes on the \emph{nehalem} cluster
at the High Performance Computing Center Stuttgart (HLRS). PL is supported by the Alexander von Humboldt Foundation.

%\bibliography{references}

\end{document}